\documentclass{article}
\usepackage{iclr2025_conference,times}

\usepackage{dblfloatfix}
\usepackage{caption}
\usepackage{booktabs}
\usepackage{overpic}
\usepackage{wrapfig}
\usepackage{subcaption}
\usepackage{listings}
\usepackage{longtable}
\usepackage{dramatist}
\usepackage{xspace}
\usepackage{pifont}
\usepackage{multirow}
\usepackage{tcolorbox}
\usepackage{xltabular}
\usepackage{longtable}
\usepackage{hyperref}
\usepackage{tikz}
\usepackage{datetime}
\usepackage{printlen}

\usepackage{amsfonts}
\usepackage{amsmath}
\usepackage{amssymb}
\usepackage{lineno}
\usepackage{sidecap}

\usepackage[bottom]{footmisc}

\title{Dargana: fine-tuning EarthPT for dynamic\\ tree canopy mapping from space\\}

\author{\hspace{-0.75em}\begin{tabular}{l}
Michael J. Smith \quad Luke Fleming \quad James E. Geach \quad Ryan J. Roberts \\
{\normalfont\texttt{\{mike.smith,luke.fleming,james.geach,ryan.roberts\}@aspiaspace.com}} \\[4pt]
Freddie Kalaitzis \quad James Banister \\
{\normalfont\texttt{\{freddie.kalaitzis,james.banister\}@aspiaspace.com}}\\[4pt]
{\normalfont Aspia Space}\\[-10pt]
\end{tabular}}

\iclrfinalcopy

\begin{document}

\maketitle

\begin{abstract}
    We present Dargana, a fine-tuned variant of the EarthPT time-series foundation model that achieves specialisation using $<$3\% of its pre-training data volume and 5\% of its pre-training compute. 
    Dargana is fine-tuned to generate regularly updated classification of tree canopy cover at 10\,m resolution, distinguishing conifer and broadleaved tree types. Using Cornwall, UK, as a test case, the model achieves a pixel-level ROC-AUC of 0.98 and a PR-AUC of 0.83 on unseen satellite imagery. Dargana can identify fine structures like hedgerows and coppice below the training sample limit, and can track temporal changes to canopy cover such as new woodland establishment. Our results demonstrate how pre-trained Large Observation Models like EarthPT can be specialised for granular, dynamic land cover monitoring from space, providing a valuable, scalable tool for natural capital management and conservation.
\end{abstract}

\vspace{-0.5em}
\begin{figure}[h]
        \begin{overpic}[width=0.6\textwidth]{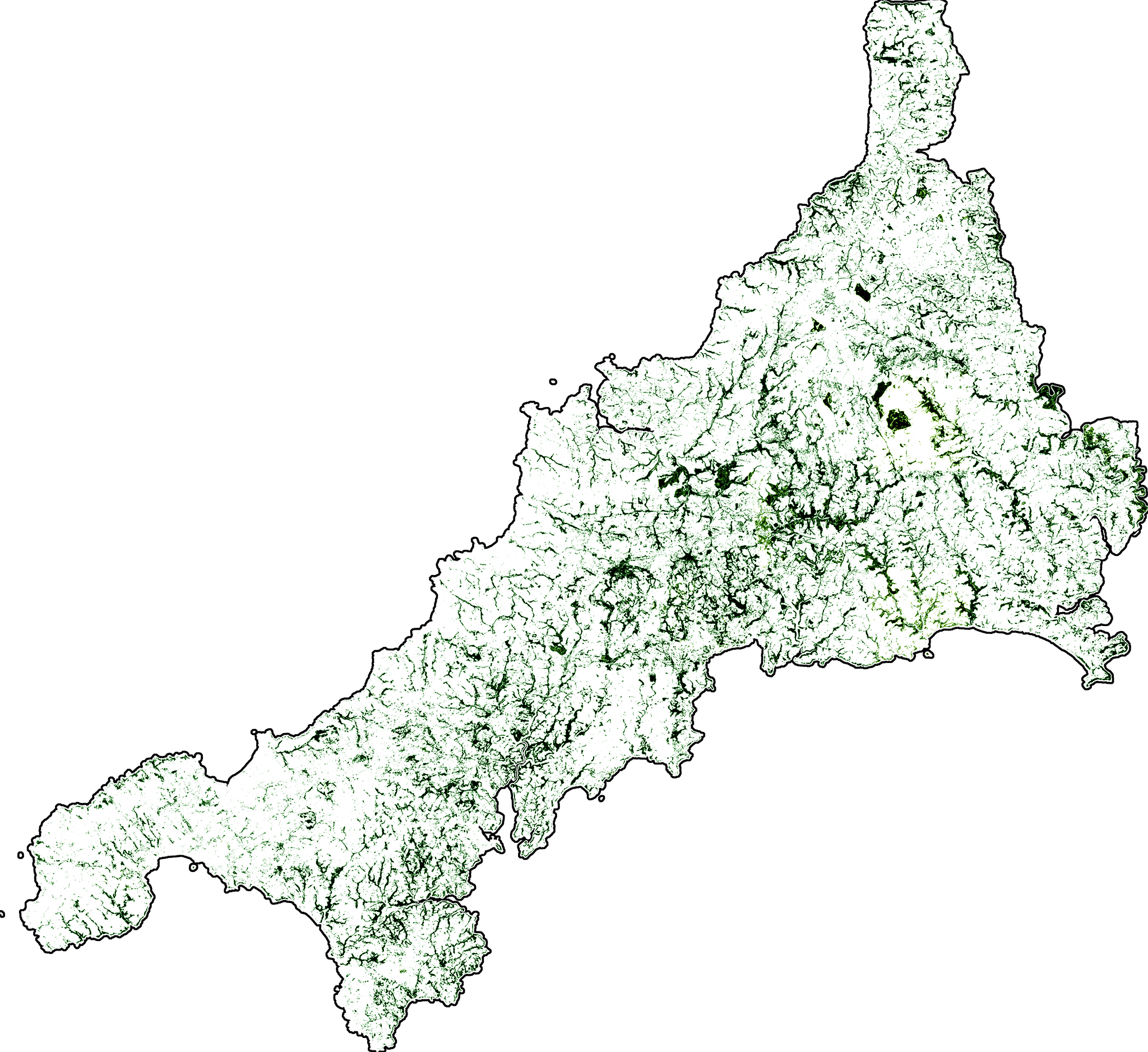}
        \put(63,15){\parbox{0.6\textwidth}{
            Figure \ref*{fig_cornwall}: Shaded pixels indicate where Dargana identified the presence of trees across Cornwall, UK on 31st December 2023.
            A higher resolution version of this map is shown in the Appendix (Fig.~\ref{fig_bigcornwall}). 
        }}
    \end{overpic}
    \vspace{-\baselineskip}
    \captionsetup{labelformat=empty}
    \caption{}
    \label{fig_cornwall}
\end{figure}
\vspace{-2em}

\section{Introduction} \label{sec_intro}

Foundation models, trained on myriad unlabelled data, have revolutionised machine learning across multiple domains \citep{ref_cong2022,ref_touvron2023,ref_jakubik2023,ref_smith2024,ref_deepseek2025}.
Within the field of Earth Observation (EO)\footnote{
    And beyond: abundant data from the observational sciences could be one solution to deep learning's `token crisis' \citep{ref_villalobos2022,ref_smith2023_aem,ref_xue2023}.
}, these models hold particular promise due to the unprecedented volume of globally consistent data across a wide range of distinct but complementary sensors, spanning synthetic aperture radar (SAR), thermal infrared and multi- and hyper-spectral optical imagery.
Previously, we introduced EarthPT---a Large Observation Model (LOM) trained on time sequences of optical and SAR imagery---demonstrating that the model could encode rich representations of the dynamic observational properties of the Earth's surface \citep{ref_smith2023}.
However, a challenge remains in efficiently specialising LOMs for specific and useful downstream tasks.

Recent work has demonstrated the potential of foundation models for high-resolution canopy mapping.
Notably, \citet{ref_tolan2024} introduced a vision transformer-based approach \citep{ref_oquab2023} for the estimation of canopy height from RGB imagery.
While this approach excels at fine spatial detail, the model operates on single time stamps and focuses solely on canopy height estimation (with tree classification being a useful by-product of their model).
In contrast, we present Dargana, a specialised variant of EarthPT that leverages the temporal capabilities of its parent model to produce continuous near-real-time tree canopy mapping and classification. 
By leveraging representations learned during pre-training on 30\,billion $8\times8$ pixel patches of multi-spectral optical and SAR observations, Dargana achieves a high level of performance in tree classification, identifying not just the presence of trees but distinguishing between broadleaved and conifer types. 

A key innovation in our approach is Dargana's ability to operate dynamically: continuously updating class probability estimates as new observations become available. 
This is a significant advance over conventional static mapping approaches, enabling regular and reliable near-time monitoring of change, such as the establishment of new planting, felling events, and density changes associated with canopy deterioration or storm damage. 
Using the county of Cornwall, UK as our study area, we demonstrate how Dargana extrapolates below the limit of its fine-tuning data---which is derived from the National Forest Inventory (NFI)---to identify the presence of trees at granularities well below the NFI's 0.5\,hectares (ha) minimum mapping unit, even approaching the single-tree limit.
The model has significant potential for remote forest and woodland monitoring and, more generally, environmental and land cover mapping from hyper-local to continental scales.

\section{Method} \label{sec_method}

\begin{wrapfigure}[19]{R}{0.5\textwidth}
    \vspace{-2.5em}
    \centering
    \includegraphics[width=0.98\linewidth]{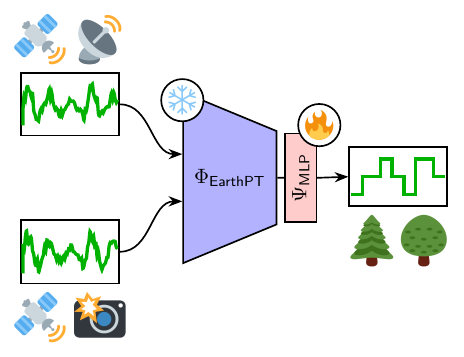}
    \caption{
        Dargana comprises of a pre-trained frozen EarthPT base model ($\Phi$) and a trainable MLP head ($\Psi$) that is appropriate for the downstream task at hand. 
        In this study that task is tree canopy type classification.
    }
\label{fig_arch}
\end{wrapfigure}
\paragraph{Pre-training and fine-tuning datasets.}

Our pre-training data consists of temporal sequences of ClearSky\footnote{
    Cloud-free multispectral optical imagery, equivalent to bottom-of-atmosphere surface reflectance values across 10 Sentinel-2 equivalent bands, spanning 400-2300\,nm.
} 
and Sentinel-1 C-band imagery of the South West of England, from 2015 to the present day. 
The imagery is split into 64 $10\times10$\,km tiles matching the British National Grid (BNG) system. 
Tiles are selected from BNG letters SW, SX and SS, excluding numbered tiles that contain no land and nine randomly sampled numbered tiles\footnote{SS21, SW86, SW94, SX08, SX29, SX36, SX38, SX46, and SX47.} that are kept unseen throughout the pre-training and fine-tuning process.
The pre-training data comprises approximately 30B $8\times8$ pixel patches of observational data spanning eight years over an area of 640\,000\,ha. 

Our fine-tuning data consists of labelled data from the NFI spanning 2018 to 2022, which has been made available under the Open Government License v3.0 by the Forestry Commission\footnote{\href{https://data-forestry.opendata.arcgis.com/}{\texttt{\color{blue}{data-forestry.opendata.arcgis.com}}}}. 
The NFI data aims to provide an inventory of all forest and woodland area over 0.5\,ha with a minimum of 20\% canopy cover, or the potential to achieve it, and a minimum width of 20\,m.
We use the NFI class labels \{{\tt broadleaved, conifer, mixed\_mainly\_broadleaved, mixed\_mainly\_conifer, young\_trees, felled, other}\} to prepare a spatio-temporal labelled training set, with `{\tt other}' representing `no tree' for our purposes. 
One caveat is that pixels labelled {\tt other} {\it will} contain trees in some cases, since the inventory is contiguous-area limited, i.e., the survey is missing canopy cover tracing (for example) hedgerows, small coppice, isolated trees, etc. 
We hypothesise that these will be a trace contaminant of the `no tree' labels given the relative covering fraction across all not-tree land cover, but this underlines the importance of high quality and granular training data. 

\paragraph{Technical approach.}

Our training routine involves first pre-training a base 300M parameter EarthPT model\footnote{See Tab.~\ref{tab_hparams} for our chosen hyperparameters and Fig.~\ref{fig_arch} for a high-level overview of our architecture.} on 30B $8\times8$ pixel patches using the same approach described in \citet{ref_smith2023}.
Once pre-trained, we freeze the base model's weights.
A small multilayer perceptron (MLP) is then attached to the central layer of the base model such that it takes as input the output of the base model's central layer, and produces as output logits suitable for the tree cover classification task.

Fine-tuning involves training the modified model on the NFI `ground truth' dataset in the typical supervised manner via the cross entropy loss.
This process is computationally cheap relative to the pre-training stage, requiring only 5\% of the compute of the pre-training step and less than 3\% of the data volume (see Tab.~\ref{tab_hparams}).
This is because we are not required to update all weights in the base model, only those in the MLP head.
At every time step, the model predicts a set of logit probabilities for every pixel, where each probability is associated with one of the class labels described above. 
The class with the highest probability is the most likely label for that pixel among known classes, however the probabilities themselves encode useful information even when a given class might not be the most likely.
We will explore this interesting property in more detail in the next section.

\section{Results} \label{sec_results}

\begin{figure}[th]
    \begin{subfigure}[b]{\textwidth}
        \includegraphics[width=\textwidth]{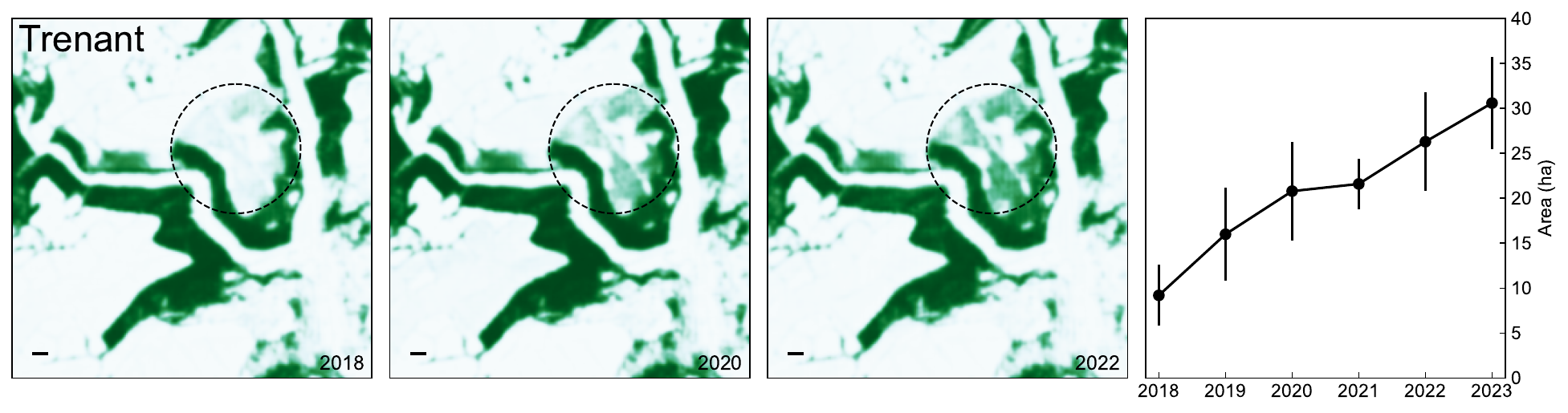}
        \caption{
            An image focused on Trenant Wood, near Looe, Cornwall, UK, with each panel showing aggregated probability for all tree classes over the calendar years 2018, 2020, and 2022. 
            The dashed circle ($r = 45$\,m) identifies a region of newly planted woodland adjacent to the ancient Trenant woodland. 
            Dargana is able to track the establishment of the new planting across the peninsula, with an increasing class probability over time tracking the increasing density of cover. 
            This demonstrates a method to quantify change in cover (and therefore remotely monitor establishment of new planting) which is shown in the right chart. 
            The area is calculated by interpreting the class probability as a covering fraction within the circular area of interest, revealing a $\sim$20\,ha increase over five years. 
            Error bars show the standard deviation of areas measured across predictions derived from individual observations across each year.
        }
        \label{fig_trenant}
    \end{subfigure}
    \begin{subfigure}[b]{\textwidth}
        \includegraphics[width=\textwidth]{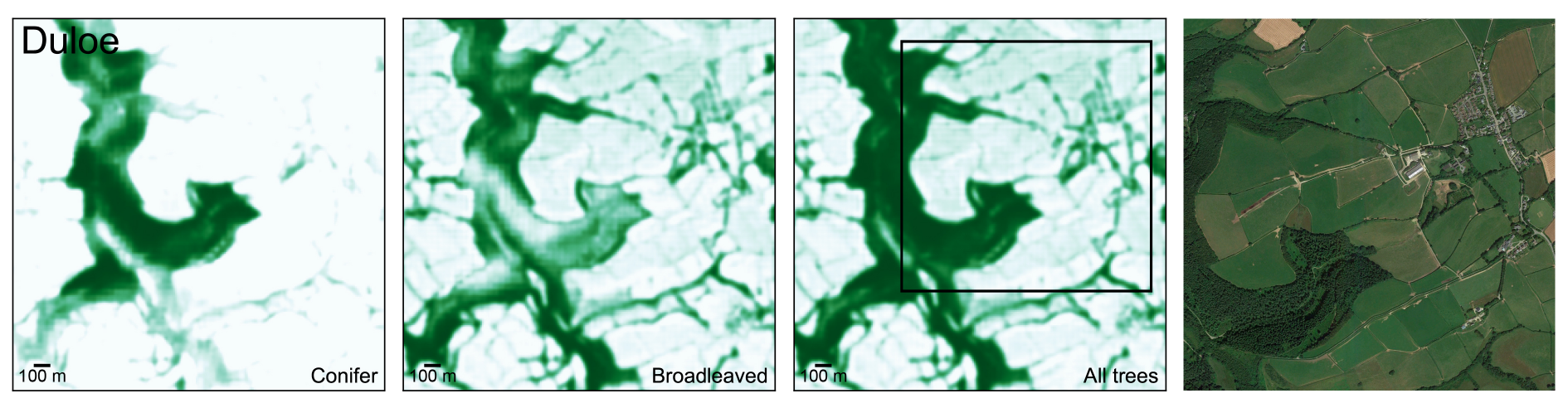}
        \caption{
            An image focused on Duloe, Cornwall, UK, with predictions of the probability of conifer, broadleaved, and both tree types.
            This region contains woodland that is a composite of managed (conifer) forestry mixed with more ancient broadleaved woodland. 
            Dargana is able to separate the tree classes within the single contiguous region of forestry.
            The two central panels illustrate how Dargana extrapolates far below the limit of the training set (for a comparison see the bottom row of Fig.~\ref{fig_sx25}), predicting the presence of broadleaved trees tracing hedgerows, coppice, trees within built-up environment and even---at the limit of the model---large lone trees. 
            Gaps and low cover within areas of contiguous woodland are also identified. 
            For visual reference we show a 2021 Google Earth view of an inset region.
        }
        \label{fig_duloe}
    \end{subfigure}
    \caption{
        Case studies showing  (a): evidence of Dargana's ability to track change and (b): information extraction below the spatial limit of the fine-tuning dataset.
    }
    \label{fig_casestudies}
\end{figure}

For testing, validation and demonstration purposes we have used the model to infer three sample datasets: \textbf{(1)} A complete mosaic of class predictions for Cornwall on 2023-12-31 (Fig.~\ref{fig_cornwall}); \textbf{(2)} A complete time series of class predictions for the unseen SX36 tile from 2018 to 2023; \textbf{(3)} A complete time series of class predictions for the case-study SX25 tile from 2018 to 2023. Figure~\ref{fig_casestudies} presents two case studies of areas within the BNG tile SX25, and a complete view of SX25 is shown in the Appendix Fig.~\ref{fig_sx25}, which includes a comparison to the NFI reference fine-tuning data. 

Our headline figure (Fig.~\ref{fig_cornwall}) presents a full mosaic of Cornwall on the 31st December 2023. 
For illustration purposes, we combine all `tree-like' classes\footnote{
\{{\tt broadleaved,\allowbreak conifer,\allowbreak mixed\_mainly\_broadleaved,\allowbreak mixed\_mainly\_conifer,\allowbreak young\_trees\}}}
into a single class. In Fig.~\ref{fig_trenant} we show an area surrounding the peninsula formed by the convergence of the East and West Looe rivers where recent tree planting is known to have occurred. 
This region is a good example of the dynamic nature of the model in its ability to revise class label probability as new observations are taken. We show annual aggregated `tree-like' probability maps from 2018, 2020, and 2022, clearly showing the progressive establishment of the young trees via the increasing class probability which appears to be serving as a proxy for canopy density. 
These predictions can be used to quantify change in tree cover, either by counting class labels or treating probabilities as a covering fraction per pixel. Fig.~\ref{fig_duloe} focuses on an area around the village of Duloe, which contains an example of woodland that is a composite of managed conifer forestry amongst more ancient broadleaved woodland, along the valley of the West Looe River (see also Appendix Fig.~\ref{fig_sx25}). Dargana cleanly separates pixels belonging to the two tree type classes within the same contiguous woodland. 
We can also see evidence here for classification of trees below the area limit of our fine-tuning dataset, with broadleaved trees tracing hedgerows (and even large isolated trees) becoming visible in the probability maps, low-density regions within the contiguous woodland, and the identification of trees within built environments.

\begin{wraptable}[11]{R}{0.35\textwidth}
    \centering
    \caption{
        Area under curve metrics for our test `tree' vs `not-tree' classification task on the SX36 tile.
    }
    \label{tab_roc}
    \begin{tabular}{lc}
        \toprule
        \textbf{Curve} & \textbf{AUC} \\
        \midrule
        ROC & 0.98 \\
        Precision-Recall & 0.83 \\
        \bottomrule
    \end{tabular}
\end{wraptable}
We quantify overall performance using the representative but unseen SX36 tile, comparing class predictions (at a threshold probability) to the NFI ground truth. 
Table~\ref{tab_roc} shows the area under curve (AUC) values calculated from the Receiver Operating Characteristic (ROC) and Precision-Recall (PR) curves measured for a tree/not-tree binary classification task over the SX36 tile, with the latter a more representative measure of performance given the unbalanced nature of the dataset, which has a relative overabundance of not-tree pixels.
Plots of the ROC and PR curves are shown in the Appendix Fig.~\ref{fig_roc}.
We consider this a good level of performance for this proof-of-concept.

\section{Conclusions} \label{sec_conclusions}

We have demonstrated that the EarthPT LOM can be specialised for dynamic---that is, regularly updated `near-time'---environmental monitoring tasks through targeted fine-tuning.  
The model's ability to track temporal change and distinguish tree types opens new possibilities for automated forest and woodland monitoring and management. 
This method is the proof-of-concept for a more generalised approach for monitoring land cover change: our results suggest that the key limitation for expanding this approach is not computational or architectural, but rather the availability of high-quality labelled training data for specific downstream tasks. 
For example, as more labelled `ground truth' data become available it would be trivial to increase the vocabulary of land cover types beyond tree-like labels to include other natural capital, surface features and even distinct events of natural or anthropogenic origin. 

The scalability of our approach is particularly promising: having demonstrated success at county scale, $\mathcal{O}(\text{100\,km})$, there are no fundamental barriers to extending coverage nationally or internationally as the underlying EarthPT architecture is inherently scalable to larger geographical regions. 
Future work will focus on expanding both the geographical scope and the range of downstream tasks, working toward a comprehensive suite of LOM-driven environmental and land-change monitoring tools built on EarthPT.

\subsubsection*{Acknowledgements}

This study made use of data compiled by the Forestry Commission, which is licensed under the Open Government License v3.0.
We thank the Good Growth Hub Grants Programme for funding this project.

\bibliography{main}

\begin{thebibliography}{13}
\providecommand{\natexlab}[1]{#1}
\providecommand{\url}[1]{\texttt{#1}}
\expandafter\ifx\csname urlstyle\endcsname\relax
  \providecommand{\doi}[1]{doi: #1}\else
  \providecommand{\doi}{doi: \begingroup \urlstyle{rm}\Url}\fi

\bibitem[Cong et~al.(2022)Cong, Khanna, Meng, Liu, Rozi, He, Burke, Lobell, and
  Ermon]{ref_cong2022}
Y.~Cong, S.~Khanna, C.~Meng, P.~Liu, E.~Rozi, Y.~He, M.~Burke, D.~Lobell, and
  S.~Ermon.
\newblock {SatMAE: Pre-training Transformers for Temporal and Multi-Spectral
  Satellite Imagery}.
\newblock \emph{Advances in Neural Information Processing Systems},
  35:\penalty0 197--211, 2022.

\bibitem[{DeepSeek AI}(2025)]{ref_deepseek2025}
{DeepSeek AI}.
\newblock {DeepSeek-R1: Incentivizing Reasoning Capability in LLMs via
  Reinforcement Learning}.
\newblock \emph{ArXiv e-prints}, 2025.
\newblock \doi{10.48550/arXiv.2501.12948}.

\bibitem[Hendrycks \& Gimpel(2016)Hendrycks and Gimpel]{ref_hendrycks2016}
D.~Hendrycks and K.~Gimpel.
\newblock {Gaussian Error Linear Units (GELUs)}.
\newblock \emph{ArXiv e-prints}, 2016.
\newblock \doi{10.48550/arXiv.1606.08415}.

\bibitem[Jakubik et~al.(2023)Jakubik, Roy, Phillips, Fraccaro, Godwin,
  Zadrozny, Szwarcman, Gomes, Nyirjesy, Edwards, Kimura, Simumba, Chu,
  Mukkavilli, Lambhate, Das, Bangalore, Oliveira, Muszynski, Ankur,
  Ramasubramanian, Gurung, Khallaghi, Hanxi, Li, Cecil, Ahmadi, Kordi,
  Alemohammad, Maskey, Ganti, Weldemariam, and Ramachandran]{ref_jakubik2023}
J.~Jakubik, S.~Roy, C.~E. Phillips, P.~Fraccaro, D.~Godwin, B.~Zadrozny,
  D.~Szwarcman, C.~Gomes, G.~Nyirjesy, B.~Edwards, D.~Kimura, N.~Simumba,
  L.~Chu, S.~K. Mukkavilli, D.~Lambhate, K.~Das, R.~Bangalore, D.~Oliveira,
  M.~Muszynski, K.~Ankur, M.~Ramasubramanian, I.~Gurung, S.~Khallaghi, Hanxi,
  Li, M.~Cecil, M.~Ahmadi, F.~Kordi, H.~Alemohammad, M.~Maskey, R.~Ganti,
  K.~Weldemariam, and R.~Ramachandran.
\newblock {Foundation Models for Generalist Geospatial Artificial
  Intelligence}.
\newblock \emph{ArXiv e-prints}, 2023.
\newblock \doi{10.48550/arXiv.2310.18660}.

\bibitem[Loshchilov \& Hutter(2017)Loshchilov and Hutter]{ref_loshchilov2017}
I.~Loshchilov and F.~Hutter.
\newblock {Decoupled Weight Decay Regularization}.
\newblock \emph{ArXiv e-prints}, 2017.
\newblock \doi{10.48550/arXiv.1711.05101}.

\bibitem[Oquab et~al.(2023)Oquab, Darcet, Moutakanni, Vo, Szafraniec, Khalidov,
  Fernandez, Haziza, Massa, El-Nouby, Assran, Ballas, Galuba, Howes, Huang, Li,
  Misra, Rabbat, Sharma, Synnaeve, Xu, Jegou, Mairal, Labatut, Joulin, and
  Bojanowski]{ref_oquab2023}
M.~Oquab, T.~Darcet, T.~Moutakanni, H.~Vo, M.~Szafraniec, V.~Khalidov,
  P.~Fernandez, D.~Haziza, F.~Massa, A.~El-Nouby, M.~Assran, N.~Ballas,
  W.~Galuba, R.~Howes, P.-Y. Huang, S.-W. Li, I.~Misra, M.~Rabbat, V.~Sharma,
  G.~Synnaeve, H.~Xu, H.~Jegou, J.~Mairal, P.~Labatut, A.~Joulin, and
  P.~Bojanowski.
\newblock {DINOv2: Learning Robust Visual Features without Supervision}.
\newblock \emph{ArXiv e-prints}, 2023.
\newblock \doi{10.48550/arXiv.2304.07193}.

\bibitem[Smith \& Geach(2023)Smith and Geach]{ref_smith2023_aem}
M.~J. Smith and J.~E. Geach.
\newblock {Astronomia ex machina: a history, primer and outlook on neural
  networks in astronomy}.
\newblock \emph{Royal Society Open Science}, 10\penalty0 (5), 2023.
\newblock ISSN 2054-5703.
\newblock \doi{10.1098/rsos.221454}.

\bibitem[Smith et~al.(2023)Smith, Fleming, and Geach]{ref_smith2023}
M.~J. Smith, L.~Fleming, and J.~E. Geach.
\newblock {EarthPT: a time series foundation model for Earth Observation}.
\newblock \emph{ArXiv e-prints}, 2023.
\newblock \doi{10.48550/arXiv.2309.07207}.

\bibitem[Smith et~al.(2024)Smith, Roberts, Angeloudi, and
  Huertas-Company]{ref_smith2024}
M.~J. Smith, R.~J. Roberts, E.~Angeloudi, and M.~Huertas-Company.
\newblock {AstroPT: Scaling Large Observation Models for Astronomy}.
\newblock \emph{ArXiv e-prints}, 2024.
\newblock \doi{10.48550/arXiv.2405.14930}.

\bibitem[Tolan et~al.(2024)Tolan, Yang, Nosarzewski, Couairon, Vo, Brandt,
  Spore, Majumdar, Haziza, Vamaraju, Moutakanni, Bojanowski, Johns, White,
  Tiecke, and Couprie]{ref_tolan2024}
J.~Tolan, H.-I. Yang, B.~Nosarzewski, G.~Couairon, H.~V. Vo, J.~Brandt,
  J.~Spore, S.~Majumdar, D.~Haziza, J.~Vamaraju, T.~Moutakanni, P.~Bojanowski,
  T.~Johns, B.~White, T.~Tiecke, and C.~Couprie.
\newblock {Very high resolution canopy height maps from RGB imagery using
  self-supervised vision transformer and convolutional decoder trained on
  aerial lidar}.
\newblock \emph{Remote Sensing of Environment}, 300:\penalty0 113888, 2024.
\newblock ISSN 0034-4257.
\newblock \doi{10.1016/j.rse.2023.113888}.

\bibitem[Touvron et~al.(2023)Touvron, Lavril, Izacard, Martinet, Lachaux,
  Lacroix, Rozi{\ifmmode\grave{e}\else\`{e}\fi}re, Goyal, Hambro, Azhar,
  Rodriguez, Joulin, Grave, and Lample]{ref_touvron2023}
H.~Touvron, T.~Lavril, G.~Izacard, X.~Martinet, M.-A. Lachaux, T.~Lacroix,
  B.~Rozi{\ifmmode\grave{e}\else\`{e}\fi}re, N.~Goyal, E.~Hambro, F.~Azhar,
  A.~Rodriguez, A.~Joulin, E.~Grave, and G.~Lample.
\newblock {LLaMA: Open and Efficient Foundation Language Models}.
\newblock \emph{ArXiv e-prints}, 2023.
\newblock \doi{10.48550/arXiv.2302.13971}.

\bibitem[Villalobos et~al.(2022)Villalobos, Ho, Sevilla, Besiroglu, Heim, and
  Hobbhahn]{ref_villalobos2022}
P.~Villalobos, A.~Ho, J.~Sevilla, T.~Besiroglu, L.~Heim, and M.~Hobbhahn.
\newblock {Will we run out of data? Limits of LLM scaling based on
  human-generated data}.
\newblock \emph{ArXiv e-prints}, 2022.
\newblock \doi{10.48550/arXiv.2211.04325}.

\bibitem[Xue et~al.(2023)Xue, Fu, Zhou, Zheng, and You]{ref_xue2023}
F.~Xue, Y.~Fu, W.~Zhou, Z.~Zheng, and Y.~You.
\newblock To repeat or not to repeat: insights from scaling llm under
  token-crisis.
\newblock In \emph{Proceedings of the 37th International Conference on Neural
  Information Processing Systems}, NIPS '23, Red Hook, NY, USA, 2023. Curran
  Associates Inc.

\end{thebibliography}
\bibliographystyle{iclr2025_conference}

\clearpage

\appendix

\section{Hyperparameters}

\begin{table*}[h!t]
\centering
\caption{Hyperparameters used in pre-training EarthPT and fine-tuning Dargana. The underlying EarthPT model is trained on approximately 30B observation tokens across 640,000\,ha of the South West of England. Fine-tuning is performed on the NFI dataset for Cornwall.}
\label{tab_hparams}
\begin{tabular}{lll}
\toprule
\textbf{Parameter} & \textbf{EarthPT Pre-training} & \textbf{Dargana Fine-tuning} \\
\midrule
\multicolumn{3}{l}{\textit{Model Architecture}} \\
\midrule
Active parameters & 300M & 150M \\
Number of layers & 24 & 14 \\
Model dimension & 1024 & 1024 \\
Number of heads & 16 & 16 \\
Block size & 256 & 256 \\
\midrule
\multicolumn{3}{l}{\textit{Training Configuration}} \\
\midrule
Batch size & 327\,680 & 163\,840 \\
Learning rate & $2\times10^{-4}$ & $2\times10^{-5}$ \\
Weight decay & 0.1 & 0.1 \\
Warmup steps & 2000 & 500 \\
Training steps & 90\,000 & 5000 \\
Optimiser & AdamW & AdamW \citep{ref_loshchilov2017} \\
\midrule
\multicolumn{3}{l}{\textit{MLP Head (Fine-tuning only)}} \\
\midrule
Hidden layers & -- & 1 \\
Hidden dimension & -- & 4096 \\
Activation & -- & GELU \citep{ref_hendrycks2016} \\
Output classes & -- & 23 \\
\bottomrule
\end{tabular}
\end{table*}

\clearpage

\section{Additional results}

\vfill
\begin{figure}[htbp]
    \begin{overpic}[width=\textwidth,angle=90]{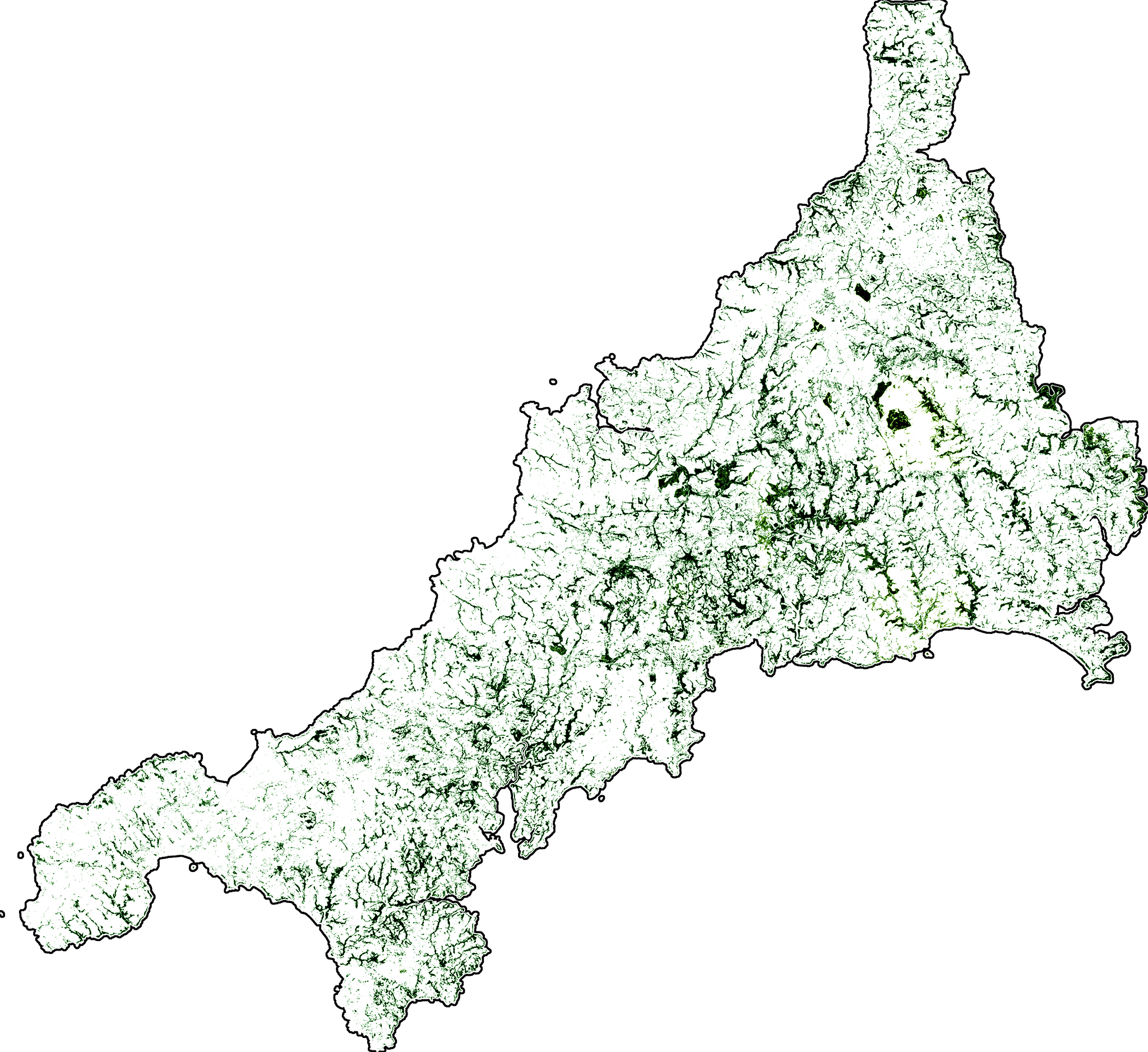}
        \put(70,60){\rotatebox{90}{\parbox{0.18\paperheight}{
            Figure \ref*{fig_bigcornwall}: A larger copy of Fig.~\ref{fig_cornwall} showing regions classified as dominated by trees, as predicted by Dargana across Cornwall, UK on 31st December 2023.
        }}}
    \end{overpic}
    \vspace{-\baselineskip}
    \captionsetup{labelformat=empty}
    \caption{}
    \label{fig_bigcornwall}
\end{figure}
\vfill

\begin{figure}[htbp]
\centering
\includegraphics[width=\textwidth]{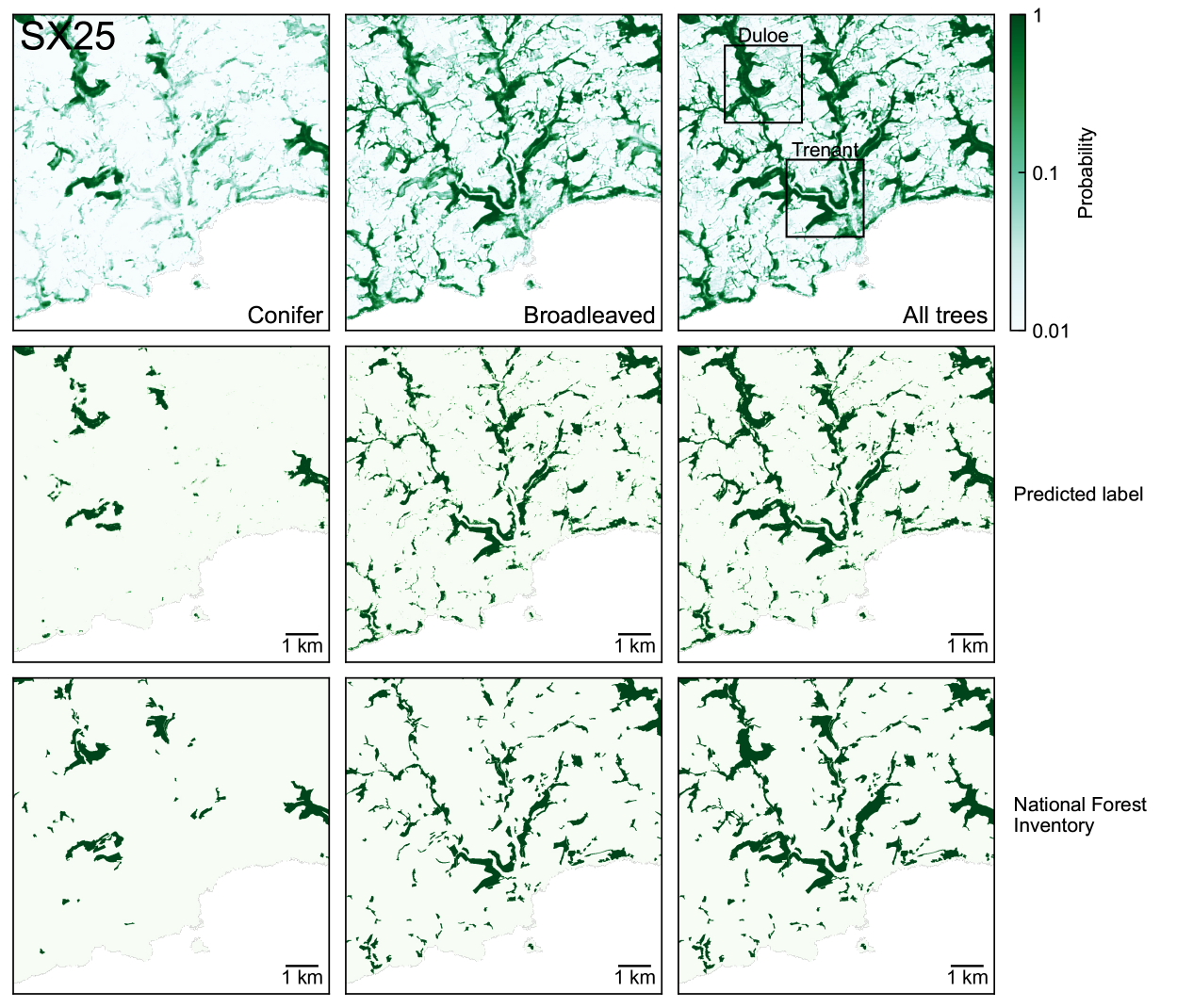}
\caption{
    Class probabilities for conifer and broadleaved tree types for SX25. 
    (Top) pixels are coloured according to class probability on a logarithmic scaling to emphasise fine structure; (middle) pixels are coloured green where the probability was the maximum across all possible classes; (bottom) National Forest Inventory labels. 
    The right column shows the combination of conifer and broadleaved tree types. 
    Note how Dargana identifies structure not present in the training data.
    We identify two sub regions `Trenant' and `Duloe' that are examined as case-studies in Fig.\,\ref{fig_casestudies}.
}
\label{fig_sx25}
\end{figure}

\begin{figure}[htbp]
    \centering
    \includegraphics[width=0.45\textwidth]{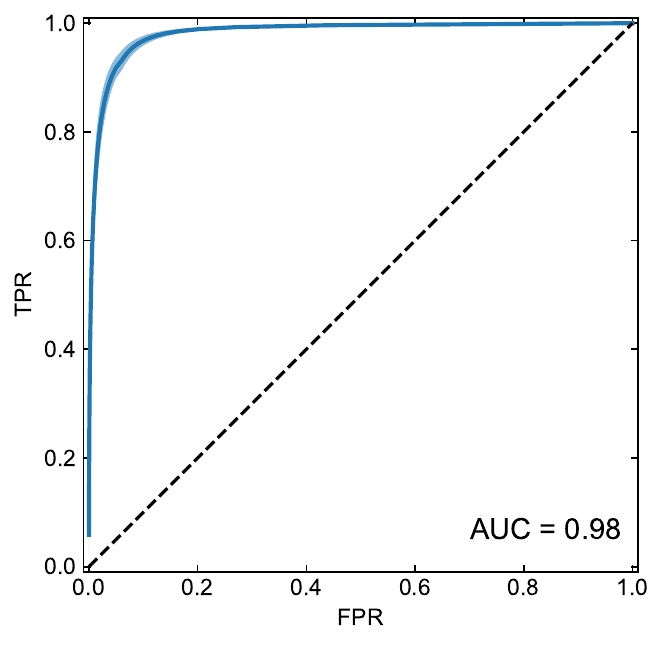}
    \hspace{1em}
    \includegraphics[width=0.45\textwidth]{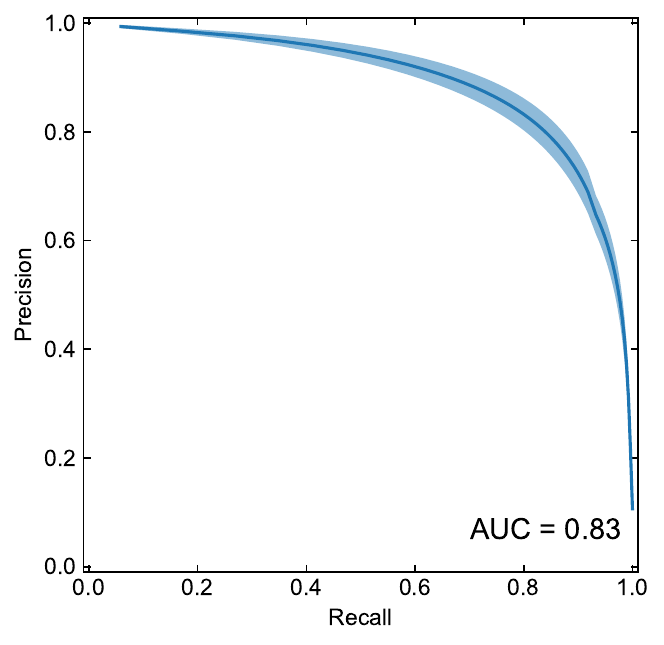}
    \caption{Left: ROC curve for the tree/not-tree binary classification task on SX36 described in \S\ref{sec_results}. Right: Precision-recall curve for the same task.}
    \label{fig_roc}
\end{figure}

\end{document}